\documentclass[11pt,twoside]{article}
 
\usepackage{asp2006}
\usepackage{graphicx}

\begin{document}
\title{Galaxy Mass Growth in GDDS and SDSS}   
\author{Karl Glazebrook$^1$, Erin Mentuch$^2$, Pat McCarthy$^3$ and Roberto Abraham$^2$ (for GDDS), Ivan Baldry$^4$ and Simon Driver$^5$ (for SDSS)}   
\affil{(1) Swinburne University of Technology, (2) University of Toronto, (3) Carnegie Observatories, (4) Liverpool John Moores University, (5) St Andrews University}    

\begin{abstract} 

I present some new results related to our understanding of the masses of galaxies both in the
local and high-redshift Universe. At high-redshift new Spitzer data on galaxies in the Gemini Deep Deep Survey
allow us a more accurate measure of stellar mass to light ratios (using rest frame near-IR light) showing a refinement of the 
measurements but not great discrepancies. In the local universe a new method is explored to estimate the {\it baryonic} mass 
function of galaxies including contributions from unseen HI. This points to an interesting result: that the baryonic mass function of galaxies may in fact be quite steep, of comparable slope to the mass function of dark matter haloes.

\end{abstract}



\section{Introduction}

The mass assembly history of galaxies is an important probe of models of galaxy formation and evolution. In recent years near-infrared selected surveys have allowed measurements of this to high-redshift (Fontana et al, 2003, 2006, Glazebrook et al. 2004). In particular the main aspect probed is the {\em stellar mass}, as beyond the very nearby Universe most galaxies are only revealed by their optical starlight emission. The nature of the game is to use multi-colour data to determine the stellar populations of distant galaxies and hence their mass-to-light ratio. There is of course uncertainty in this determination and to address it this is best done at redder wavelengths ($\lambda \ga 0.8\micron$) where mass-to-light ratios vary less. This is due to the dominance of the near-IR light by older long-lived stellar populations. In contrast as one goes bluer ($\lambda < 0.4\micron$) the variation increases dramatically. In the rest-frame UV the range is many orders of magnitudes reflecting the fact that these wavelengths really measure {\em star formation rate} rather than stellar mass. Infrared $K$-band surveys such as 2MASS have allowed very accurate measurements of the local stellar mass function (e.g. Cole et al. 2001). 

\section{Galaxy Masses in GDDS}

The Gemini Deep Deep Survey (GDDS) was an infrared selected ($K<20.6$) redshift survey designed to be color-complete to $z=2$ (Abraham et al. 2004). The galaxy mass density at $1<z<2$ (Glazebrook et al. 2004, GDDS Paper III) showed a small drop in the number of massive galaxies compared to $z=0$ but not the large drop expected in models where stellar mass assembly tracked cold dark matter.

\begin{figure}[t]
\begin{center}
\includegraphics[width=6cm, angle = -90]{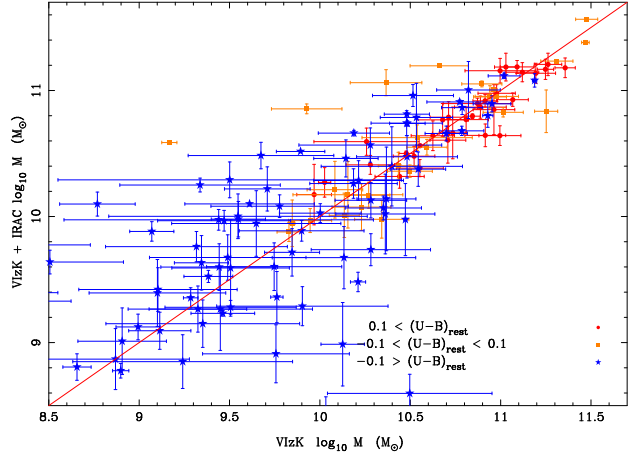}
\end{center}
\caption{Comparison of masses obtained by fitting to VIzK photometry and VIzK + IRAC channels 1,2,3,4 (3,4,5,8\micron) The legend shows the break down by rest frame $U-B$ colour (as interpolated from the SED fitting). The star-formation histories are fitted with a two-component model (including bursts). 
}\label{fig1}
\end{figure}

One limitation of the GDDS was that at the highest redshifts the observed $K$-band data was only probing rest-frame $R$-band, where the mass-to-light ratio might not be as robust as we might like. For example Shapley et al (2005) using Spitzer data for a $z\sim 2$ sample found that even the rest-frame $H$-band mass-to-light varied by up to a factor of 15 highlighting the importance of obtaining the very broadest band SED data. Note though her sample was dominated by young star-forming galaxies which naturally show the greatest variation.

Accordingly in Cycle 1 we obtained Spitzer IRAC imaging of several GDDS fields to extend the spectral energy distribution (SED) coverage to the 3--8\micron\ bands. In Mentuch at el (2008, in prep.) we have extended the very general 2-population mass-fitting methods of Paper III to include the extra data. This mass fitting code includes the effect of both simultaneous old and young (burst) stellar populations. The basic result is shown in Figure~1 which compares the stellar mass estimates with and without the IRAC data. 

We can sort the galaxies by the rest-frame $U-B$ colour as determined by the same SED fitting. As expected from a near-IR selected sample most of the more massive galaxies are redder and at low masses we are dominated by young blue objects. On average we find negligible shift in the mean stellar mass from including the IRAC data --- galaxies scatter along the 1:1 line. For the red objects the scatter is smaller (0.14 dex). This is as expected as redder more massive objects are dominated by older stellar populations and any recent star-formation with consequent effects on the SED is relatively small. For the lowest mass blue objects there is much more scatter (0.59 dex) --- and individual mass estimates can be wrong by up to a factor of ten! This accords with the findings of Shapley et al. Encouragingly though the mean trend is still quite robust which we attribute to the general nature of the population models which marginalize the mass over a very diverse range of possible star-formation histories.

Further we find that only the IRAC channels 1+2 are really required and adding channels 3+4 adds little mass information. We also find little redshift dependence in the mass offsets. Using revised IRAC masses we have recomputed the mass assembly following Paper III and find no significant change. The mass density at $1.2<z<2$ is $\sim 1.5\times$ higher using the IRAC masses but this is within the previous error bars. We will present this result and further ongoing work, including the effect of using the Maraston (2005) population synthesis models (which have an alternate TP-AGB star treatment, critical for old populations), in Mentuch et al. 2008.

\section{Galaxy Masses in SDSS}

The true galactic baryonic mass function (GBMF)  of the local universe is an important quantity to determine, it measures the fraction of cosmic baryons which have collapsed to form galaxies (Read \& Trentham 2005). The stellar mass component (GSMF) is now robustly determined by SDSS and 2MASS but it is well know that as one proceeds to low-mass galaxies the mass is increasingly dominated by cold HI gas (e.g. Swaters \& Bahcells 2002). Unfortunately direct determination of this HI content is not possible over representative cosmic volumes such as that probed by SDSS. 

We have developed a new approach to determining the GBMF from SDSS (Baldry, Glazebrook \& Driver, 2008). The key assumption here is that the stellar mass-metallicity relationship (MRZ; which is a tight correlation well determined by SDSS, e.g. Tremonti et al. 2004) can be used as a probe of unseen cold gas baryons. In a closed box model as cold gas is processed in to stars the metallicity increases as the fraction processed increases. In BGD08 we show this is quite robust against small deviations from the closed box model. Using our interpretation of the MZR, which is a critical assumption of this method, we can then work out the gas/star ratio as a function of stellar mass and compute a GBMF averaged over the entire SDSS survey. The full details are presented in BGD08. The main steps are:

\begin{enumerate}
\item Compute stellar masses from the $ugriz$ photometry of the SDSS sample. We use 49,968 galaxies from the NYU-VAGC catalog (Blanton et all. 2005) with $z<0.05$ and the redshifts have been corrected for peculiar velocities using a flow model. The mass fitting follows the methods of Glazebrook et a. 2004 and agrees well with other authors.
\item We compute the GSMF using a methodology which corrects for large-scale structure effects. 
\item We parameterize the MZR of Tremonti et al. and fit it with a 3 component (outflows, stars, gas mass) model following Edmunds (1990). In the end we find the effect of outflows are negligible and set this to zero resulting in a closed box model.
\item We use the fits to convert stellar masses to baryonic masses as a function of mass and determine the GBMF (Figure~2).
\end{enumerate}

\begin{figure}[t]
\begin{center}
\includegraphics[width=12cm]{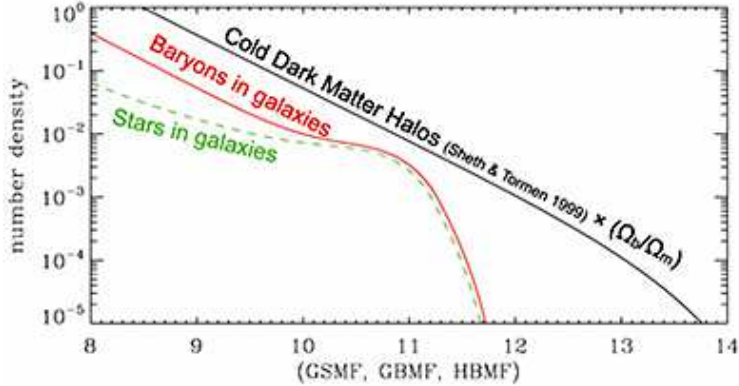}
\end{center}
\caption{Comparison of SDSS galaxy mass functions in stars (GSMF) and baryons (GBMF) following our procedure. For comparison the theoretical mass function of cold dark matter haloes of Sheth \& Tormen 1999) is shown, this is scaled by the cosmological baryon fraction to make a `Halo Baryonic Mass Function (HBMF)'.  
}\label{fig2}
\end{figure}

We find that the GBMF is much steeper at the low mass end, this is because of the increasing gas:stars ratio. As a sanity check we have compared these ratios to those directly measured in the nearby Universe by HI surveys, we find reasonable agreement as discussed in BGD08, which encourages us that our methodology is correct. Remarkably we find that the slope of the low-mass end is the same as that for dark matter haloes in cosmological simulations ($dN/dM \propto M^{-1.9}$). Historically the `flatness of the galaxy luminosity function' has been something of a problem. We find it is is not flat, i.e. there is no `small scale structure problem' (Moore et al. 1999) once all the mass is accounted for, at least down to SMC masses. Alternatively one can say that Figure~2 shows that `galaxy formation efficiency' (i.e. fraction of halo baryons collapsed in to cold gas and stars) is constant at low masses at $\sim 10\%$. This may present a simplified target for models of feedback effects in galaxy formation.

\acknowledgements

This paper is based on observations obtained at the Gemini
Observatory, the NASA Spitzer Space Telescope, and the Sloan Digital Sky Survey. We acknowledge funding through Spitzer grant  RSA 1265400.



\section*{References}
\hbox{\begin{minipage}{6cm}
\footnotesize\parindent=0pt

Abraham R. G., et al., 2004, AJ, 127,  2455

Baldry I. K., Glazebrook K., Driver S. P.,  2008, MNRAS in press, (arXiv:0804.2892) 

Blanton M. R., et al., 2005, AJ, 129,  2562 
 
Cole S., et al., 2001, MNRAS, 326,  255 

Edmunds M. G., 1990, MNRAS, 246,  678

Fontana A., et al., 2003, ApJ, 594,  L9

Fontana A., et al., 2006,  A\&A, 459,  745 
Maraston C., 2005, MNRAS, 362,  799 
Moore B.,et al.,  1999, ApJ, 524,  L19 

\end{minipage}
\quad
\begin{minipage}{6cm}
\footnotesize\parindent=0pt

Read J. I., Trentham N., 2005,  RSPTA, 363,  2693 

Shapley A. E., Steidel C. C., Erb  D. K., Reddy N. A., Adelberger K. L., Pettini M., Barmby P., Huang J.,  2005, ApJ, 626,  698 

Sheth R. K., Tormen G.,  1999, MNRAS, 308,  119 

Swaters R. A., Balcells M., 2002, A\&A, 390,  863 

Tremonti C. A., et  al., 2004, ApJ, 613,  898 

\end{minipage}
}

\end{document}